\documentclass{jpp}

\usepackage[utf8]{inputenc}
\usepackage{latexsym}
\usepackage{graphics}
\usepackage{graphicx}
\usepackage{subcaption}
\usepackage{amsmath}
\usepackage{amssymb}
\usepackage[T1]{fontenc}
\usepackage[english]{babel}
\usepackage{bm}
\usepackage{xcolor}
\usepackage{tabularx}
\usepackage{booktabs}
\newcommand{\omp}{\omega_p}
\newcommand{\ie}{\emph{i.e., }}

\newcommand{\reff}[1]{(\ref{#1})}
\newcommand{\eref}[1]{Eq.\reff{#1}}
\newcommand{\erefs}[1]{Eqs.\reff{#1}}
\newcommand{\figref}[1]{Fig.\ref{#1}}
\newcolumntype{Y}{>{\centering\arraybackslash}X}

\title{\Large\textbf{Diffusive time evolution of the Grad-Shafranov Equation for a Toroidal Plasma}}

\author{{\large Giovanni Montani$^{\,1,\,2}$, Matteo Del Prete$^{\,2}$, Nakia Carlevaro$^{\,1,\,3}$, Francesco Cianfrani$^{\,4}$}
\vspace{3mm}\\
\emph{\footnotesize $^1$ENEA, Fusion and Nuclear Safety Department, C.R. Frascati, Via E. Fermi 45, 00044 Frascati (Roma), Italy}\\
\emph{\footnotesize $^2$ Physics Department, ``Sapienza'' University of Rome, P.le Aldo Moro 5, 00185 Roma, Italy}\\
\emph{\footnotesize $^3$Consorzio RFX, Corso Stati Uniti 4, 35127 Padova, Italy}\\
\emph{\footnotesize $^4$ PIIM UMR7345, CNRS, Aix-Marseille University, Jardin du Pharo, 58 Boulevard Charles Livon, 13007 Marseille, France}\\
}
\date{}
\begin{document}
\maketitle

\begin{abstract}
We describe the evolution of a plasma equilibrium having a toroidal topology in the presence of constant electric resistivity. After outlining the main analytical properties of the solution, we illustrate its physical implications by reproducing the essential features of a scenario for the upcoming Italian experiment Divertor Tokamak Test Facility, with a good degree of accuracy. Although we find the resistive diffusion timescale to be of the order of $10^4\,$s, we observe a macroscopic change in the plasma volume on a timescale of $10^2\,$s, comparable to the foreseen duration of the plasma discharge by design. In the final part of the work, we compare our self-consistent solution to the more common Solov'ev one, and to a family of nonlinear configurations.
\end{abstract}


\section{Introduction}
The theory underlying plasma equilibrium in axial symmetry, in particular in toroidal configurations like Tokamak devices \citep{wesson}, consists of the so-called Grad-Shafranov equation (GSE) \citep{biskamp,grad,shafranov}. This equation is nothing more than the implementation of the basic magnetostatic equation (\ie steady magnetohydrodynamics (MHD) in the absence of bulk plasma velocities) to the axial symmetry, once making explicit use of the magnetic flux function as fundamental variable \citep{landau8} (see also \citet{ac86} and, for a detailed review on this topic, \citet{dini} and refs therein).

In the practice of Tokamak experiments, stationary configurations have actually a finite lifetime, both for the intrinsic finiteness of the discharge duration and for the emergence of instabilities, able to grow and then to destroy such steady profiles. Nonetheless, the physical meaning of the GSE solutions is ensured by the different timescales between their validity and the growth rates of the most common instabilities \citep{biskamp}. The basic feature of Tokamak devices is their toroidal topology, characterized by a nearly constant toroidal magnetic field and a smaller poloidal component (the ratio of the latter to the former is typically taken of the order of the torus aspect ratio $\sim1/3$ \citep{wesson}). The presence of the poloidal magnetic component, mainly due to the induced current in the plasma, implies a certain rotation of the field lines around the torus axis, which improves the stability properties of charged particles in such machines.

The role of resistive diffusion in non stationary axisymmetric Tokamak plasma has been discussed originally in the seminal paper \citet{Grad:70} \citep[see also][]{Nuhrenberg:72,Grad:74,Grad:75,Pao:76,Grad:77,Gradeta:77,Reid:79,Miller:85,strand01}. The theoretical analysis outlines that two transient processes are involved: the skin effect, {\it i.e.} the same mechanism responsible for the penetration of magnetic field lines in a solid conductor, and the nonlinear diffusion of pressure across magnetic field lines. These two processes are, in general, nonlinearly coupled and they can be disentangled by considering the two limiting cases for the generalized Ohm's law: no bulk velocity, resulting in $\vec{E}=\vec{J}/\sigma$ and associated to the first process, and no electric field, resulting in $\vec{v}\wedge \vec{B}=\vec{J}/\sigma$ and associated to the second process (here, and in the following, $\vec{E}$ and $\vec{B}$ denote the electric and magnetic fields, respectively, while $\sigma$, $\vec{J}$ and $\vec{v}$ are the plasma electric conductivity, current density and fluid velocity, respectively). Indeed, the characteristic timescales of both processes are generically longer than those of instabilities and of wave propagation.

In this paper, we investigate an analytical treatment of the equilibrium of magnetized plasma in the presence of non ideal effects. We consider the case without convection and we demonstrate how the influence of resistive diffusion due to the skin effect can be treated analytically. This is due to a technical reason, already noted in \citet{Gradeta:77}: the diffusion equations for the magnetic poloidal flux and for the toroidal current function coincide in the limit of constant resistivity and no convection (see \erefs{egs10a} and \reff{egs10b} below). This allows us to construct a consistent non-stationary equilibrium solution in which the  time dependence is only within the poloidal magnetic flux function, dubbed $\psi$, which dynamics is governed by the generalized Ohm's law. In other words, we describe a diffusion process in which all the relevant plasma quantities remain instantaneously frozen on a non-stationary magnetic field configuration. In this sense, we speak of a \emph{non-stationary} GSE. 

It is worth noting that our analytical setting differs from the traditional Solov'ev configuration, which was originally studied in \citet{solo68} and has formed the basis of most analytical studies on Tokamak plasma equilibrium properties since then. To understand this, we recall that the explicit form of the GSE depends on the choice of two arbitrary functions, namely the thermodynamic pressure and the poloidal current function, as functions of $\psi$. It turns out that the Solov'ev choice, while having the merit of simplicity and versatility, is not compatible with diffusion due to constant resistivity. Here, we outline the proper assumptions to be made in order to obtain a consistent evolutive solution. We recover a class of equilibria that was actually already considered in \citet{mccar99}, even though that work was not at all motivated by dynamical considerations and was performed in an ideal, time-independent setting.
We also remark that a linear assumption on the poloidal current with respect to $\psi$, as considered in \eref{egs12}, is of interest for machines different than Tokamaks, where the plasma region includes the central axis of symmetry (\emph{e.g.}, see \citet{proto17}), while Solov'ev-like profiles are not suitable in such geometries.

After defining the lifetime of the configuration, we outline the basic eigenvalue structure of the mathematical problem and solve the relevant equation. Then, to illustrate a real physical situation, we implement this model to a specific plasma scenario for the Italian Tokamak proposal named Divertor Tokamak Test (DTT) Facility \citep{albanese17,albanese19}. We show how our solution is able to reproduce the essential features of the $5\,$MA double-null scenario described in \citet{albanese19} with a good degree of accuracy. The reconstructed equilibrium is associated to a theoretical timescale, defined as the inverse decay rate of the magnetic flux function due to resistive diffusion, of about $10^4\,$s, while the foreseen duration for the discharge according to machine parameters is about $50\,$s. However, we spot the emergence of an effective lifetime in our model, corresponding to the loss of confinement of the plasma configuration, which we observe on a timescale of $10^2\,$s comparable with the discharge duration. It is important to remark how the obtained radial pressure profiles indicate that our model refers to low confinement states only and that the presence of a pedestal, typical of the H--mode \citep{hm1,hm2,hm3}, could significantly increase the configuration lifetime.

In our study, the GSE is self-consistently verified at all times along the plasma dynamics, with an analytical expression for the equilibrium under the limiting assumption of a constant resistivity in the plasma region. In different approaches, usually used in Tokamak numerical simulations, more realistic transport dynamics are the result of specific assumptions and the consideration of flux-averaged variables, allowing for a numerical integration of the profile evolution, once the static equilibrium is assigned. Hence, in such schemes, any compatibility condition on the source terms in the GSE is neglected. In the last part of this work, to estimate the possible discrepancies of different approaches in the present case, we consider a Solov'ev-like configuration, disregarding one of the dynamical equations (namely \eref{egs-red1}), and we use this solution to model the same double-null scenario previously considered. We find the two profiles to be in good accordance at all times up to deconfinement, arguably due to the linearity of the system. As a further test, we also perform a numerical study on a family of nonlinear scenarios, taking the Solov'ev result as reference. We find a bigger discrepancy in the damping rate of the profile, hence we suggest that larger errors could arise in nonlinear situations.

The paper is organized as follows. In Sec.\ref{sec1}, we describe the basic MHD equations which characterize the dynamics of the plasma configuration. In Sec.\ref{sec2}, the details of the considered dynamical scenario as due to resistivity are developed, outlining the analytical implications of this effect and the temporal decay of the profile. In Sec.\ref{sec3}, we solve the resulting eigenvalue problem, and use our solution to model a DTT-like double-null plasma scenario. The lifetime of the configuration and its profile are outlined. In Sec.\ref{sec4}, we use the Solov'ev-like solution to model the same plasma scenario. We also provide estimates on the error associated to a class of nonlinear scenarios. Concluding remarks follow in Sec.\ref{sec5}.

\section{Basic equations}\label{sec1}
We study a plasma confined in a magnetic field $\vec{B}$ and having negligible macroscopic motion, \ie its fluid velocity $\vec{v}$ identically vanishes. The plasma is also characterized by a finite electric conductivity $\sigma\simeq const.$ The electric field in the plasma is then provided, via the current density, according to the generalized Ohm's law
\begin{equation}
\vec{E} = \frac{1}{\sigma}\vec{J}\,.
\label{egs1}
\end{equation}
Expressing the electric field via the scalar and vector potentials ($\varphi$ and $\vec{A}$, respectively), \ie 
\begin{equation}
\vec{E} = - \nabla \varphi -  \partial _t\vec{A}/c\,,
\label{egs2}
\end{equation}
observing that $\vec{B} = \nabla \wedge \vec{A}$ and noting that in the Coulomb gauge (\ie $\nabla \cdot \vec{A} = 0$) $\vec{J} =- (c/4\pi)\Delta \vec{A}$ ($c$ is the speed of light), we can finally rewrite \eref{egs1} as follows
\begin{equation}
\partial _t\vec{A} = -c\nabla \varphi+
\frac{c^2}{4\pi\sigma}\Delta \vec{A}\,.
\label{egs3}
\end{equation}

We now consider an axial symmetry, associated to a toroidal topology by the choice of cylindrical coordinates $r$, $\phi$ and $z$, having the following ranges of variation: $R_0-a\leqslant r \leqslant R_0+a$, $0\leqslant \phi < 2\pi$. Here $R_0$ denotes the major radius of a standard Tokamak configuration, while $a$ is the minor radius (we also have $|z|\lesssim a$). The axial symmetry is implemented by requiring the independence of all the physical quantities on $\phi$.

We write the vector potential as follows 
\begin{equation}
\vec{A} = \vec{A}_p + \frac{\psi}{2\pi r}\,\hat{e}_{\phi} \, , 
\label{egs5} 
\end{equation}
where $\hat{e}_{\phi}$ is the toroidal versor and the poloidal (radial-axial) vector potential $\vec{A}_p$ is described by the relations
\begin{equation}
\nabla \cdot \vec{A}_p = 0\,,\qquad 
\nabla \wedge \vec{A}_p = B_{\phi} \equiv \frac{2}{c}\frac{I}{r}\,, 
\label{egs6}
\end{equation}
The functions $\psi = \psi (t, r, z)$ and $I = I(t, r, z)$ denote the flux function and the axial current function (in the cross section $\pi r^2$), respectively, and they are the considered dynamical degrees of freedom. 

Since the scalar electric potential gradient is poloidal in axial symmetry, we easily get the dynamics of $\psi$ from the toroidal component of \eref{egs3}, and that one of $I$ by taking the curl of the remaining poloidal components (so eliminating the gradient of $\varphi$) and by taking into account \erefs{egs5} and (\ref{egs6}). Thus, we arrive to the following two (identical) dynamical equations:
\begin{align}
\partial _t\psi &= \frac{c^2}{4\pi\sigma}\Delta ^*\psi\,  ,\label{egs10a}\\
\partial _tI &= \frac{c^2}{4\pi\sigma}\Delta^{*} I\,  ,\label{egs10b}
\end{align}
where we have defined $\Delta^*(...) \equiv r\partial_r(1/r\partial_r(...)) + \partial ^2_z(...)$. Now, the toroidal component of  the momentum conservation equation ($p$ denoting the plasma pressure), \ie
\begin{equation}
\nabla p = (\nabla\wedge\vec{B})\wedge\vec{B}/4\pi\,,
\label{egs11}
\end{equation}
reduces to the constraint 
$\partial_r\psi\partial_zI - \partial_z\psi 
\partial_rI = 0$, implying the basic 
restriction $I = I(\psi)$. Once we substitute this expression into \eref{egs10b}, the compatibility with \eref{egs10a} leads to the condition
\begin{equation}
\frac{d^2 I}{d\psi^2} |\nabla \psi|^2 = 0\;,\label{egs-red1}
\end{equation}
which is either trivially solved by $\psi=const.$, or by letting
\begin{equation}
\frac{d^2I}{d\psi^2} = 0 \;\Rightarrow\;I = A_1\psi + A_0\, , 
\label{egs12}
\end{equation}
where $A_{1,0}$ are two integration constants. Near the magnetic axis, \eref{egs12} would correspond to considering a first order expansion of $I(\psi)$, in a similar fashion as in \citet{solo68}, where analytical solutions of the GSE in the same regime are found by expanding the quantities $dp/d\psi$ and $I dI/d\psi$. In this context, however, the Solov'ev solution fails to guarantee the compatibility of the resistive system, \emph{de facto} neglecting \eref{egs10b}. In Sec.\ref{sec4}, we study this uncompatible scenario in detail, providing a comparison with the formally correct solution, which is derived in the following Section.

The poloidal components of \eref{egs11} reduce to the usual GSE, 
\begin{equation}
\Delta ^*\psi = - 16\pi^3 r^2 
\frac{dp}{d\psi} - \frac{16\pi^2}{c^2}I\frac{dI}{d\psi}\,,
\label{egs13o}
\end{equation}
in which we also implement choice (\ref{egs12}), \ie
\begin{equation}
\Delta ^*\psi = - 16\pi^3 r^2 
\frac{dp}{d\psi} - \frac{16\pi^2}{c^2}(A_1^2\psi+A_1 A_0)\,. 
\label{egs13}
\end{equation}
Finally, the mass conservation equation ($\rho$ being the plasma mass density), \ie
\begin{equation}
\partial _t\rho + \nabla \cdot(\rho \vec{v}) = 0
\label{egs14}
\end{equation}
becomes, in the present scenario, the simple relation $\rho \equiv \rho_0(r,z)$, where $\rho_0$ denotes is the time independent plasma mass density.

\section{Dynamical implications}\label{sec2}
The discussion above clarified how $\psi(t,r,z)$ is the only dynamical variable of the system, which has to obey \erefs{egs10a} and (\ref{egs13}). These two equations can be combined into a new one:
\begin{equation}
\partial_t \psi = \frac{c^2}{4\pi\sigma}\left( - 16\pi^3 r^2 
\frac{dp}{d\psi} - \frac{16\pi^2}{c^2}(A_1^2\psi+A_1 A_0) \right)\,,
\label{egs15}
\end{equation}
which remains coupled to \eref{egs13}.

In order to look for analytical solutions, having to deal with a $dp/d\psi$ term, we preserve the linearity of the system by assuming the following expression for the pressure:
\begin{equation}
p(\psi) = C_2 \psi^2 /2 + C_1 \psi + C_0\,,
\label{egn1}
\end{equation}
where $C_{2,1,0}$ are generic real constants. Taking (\ref{egn1}) into account, it is easy to check that the general solution of \eref{egs15} takes the form
\begin{equation}
\psi (t,r,z) = \psi _0(r,z) e^{ - \gamma(r) t}+ \delta (r)\, , 
\label{egs18r}
\end{equation}
where $\psi_0(r,z)$ is a generic function yet to be determined, while the quantities $\gamma(r)$ and $\delta(r)$ are given by: 
\begin{equation}
\gamma(r) \equiv \frac{4\pi}{\sigma} \left(A_1^2+\pi c^2 C_2 r^2 \right) \,,\qquad
\delta (r) \equiv -\frac{\pi c^2C_1 r^2 + A_1A_0}{\pi c^2 C_2 r^2 + A_1^2} \, .
\label{egs19r}
\end{equation}
By substituting \erefs{egs18r} and (\ref{egs19r}) into \eref{egs13}, we obtain an equation for $\psi_0(r,z)$:
\begin{align}
e^{-\gamma(r)t} \left[ \Delta ^*\psi_0 +\frac{8\pi^2c^2}{\sigma} C_2 r t \left( -2\partial_r\psi_0 + \frac{8\pi^2c^2}{\sigma}C_2 r t \psi_0 \right) \right]+\frac{8\pi^2 c^4 A_1 C_2 (A_1C_1 - A_0C_2)}{(\pi c^2 C_2 r^2+A_1^2)^3} \nonumber \\
= -16\pi^3 r^2\left[ C_2 (\psi_0 e^{-\gamma(r)t}+\delta(r))+C_1 \right] - \frac{16\pi^2}{c^2}\left[ A_1^2 (\psi_0 e^{-\gamma(r)t}+\delta(r))+A_1A_0 \right] \,,
\end{align}
which involves terms proportional to $1$, $t$ and $t^2$, which of course must be equated separately. In order to solve the time dependent equations, the constant $C_2$ must be set to 0, which implies that also the pressure must be linear in the flux function, like the axial current function $I$. This kind of choice for the functions $p$ and $I$ has been previously studied in \citet{mccar99}, although in our work we show how such an assumption is naturally motivated by dynamical considerations.

After choosing $C_2=0$, \erefs{egs18r} and (\ref{egs19r}) reduce to:
\begin{equation}
\psi (t,r,z) = \psi _0(r,z) e^{ - \gamma t}+ \delta (r)\, , 
\label{egs18}
\end{equation}
and
\begin{equation}
\gamma \equiv \frac{4\pi A_1^2}{\sigma}\,,\qquad
\delta (r) \equiv -\frac{\pi c^2C_1}{A_1^2}r^2- \frac{A_0}{A_1}\, ,
\label{egs19}
\end{equation}
respectively. Finally, the equation for $\psi_0(r,z)$ takes the form:
\begin{equation}
\Delta^* \psi _0 = -16\pi ^2 A_1^2 \psi _0/c^2\, . 
\label{egs20}
\end{equation}
Before studying the morphology of the plasma profile, we observe that the magnetic configuration is always damped in time by a constant rate $\gamma$, which we associate with a resistive diffusion timescale, towards an asymptotic constant field $\vec{B}_{\infty} = (c^2C_1/A_1^2)\hat{e}_z\,$.

The present study has the merit of defining quantitatively a lifetime for a given plasma configuration, once resistive diffusion is consistently taken into account. In particular, we showed how the lifetime is very sensitive to $A_1$, \ie the proportionality constant between $I$ and $\psi$. This approach in not intended as an alternative choice to standard transport studies on assigned equilibria. In fact, we simply clarify the influence of the considered correction to Ohm's law on the evolution of a plasma profile, which could play a significant role in the physics of future steady-state Tokamak machines.

\section{Magnetic profile}\label{sec3}
In order to investigate the constant poloidal flux function $\psi_0(r,z)$ predicted by \eref{egs20}, we observe that its linearity allows to consider the following Fourier expansion:
\begin{equation}
\psi_0(r,z) = \int^{\infty}_{0} dk \chi_k(r) e^{ikz}+c.c.\,, 
\label{egs22}
\end{equation}
where $\chi_k(r)$ verifies the eigenvalue problem 
\begin{equation}
\frac{d^2\chi_k}{dr^2} - \frac{1}{r} \frac{d\chi_k}{dr} = E_k \chi_k\,, \quad
E_k \equiv k^2 - \frac{16\pi^2A_1^2}{c^2}\,.
\label{egs23}
\end{equation}
The equation for $\chi_k$ admits an analytical solution in terms of Bessel functions. In particular, defining $x\equiv r |E_k|^{1/2}$ and setting $\chi_k\equiv r\varepsilon(k,x)$, \eref{egs23} can be rewritten as 
\begin{equation}
x^2\frac{d^2\varepsilon}{dx^2} + x\frac{d\varepsilon}{dx} - \left( 1 \pm x^2\right)\varepsilon = 0\,, 
\label{egsx}
\end{equation}
where the sign $-$ corresponds to $E_k<0$, \ie to $k<k^*$, where $k^*=4\pi A_1/c$, while the sign $+$ to the case $E_k>0$, \ie to $k>k^*$.

In correspondence to the sign  $\mp$, the solutions of \eref{egsx} are
\begin{align}
\varepsilon _-(k,x) &= \varepsilon _1(k) J_1(x) + \varepsilon _2(k) Y_1(x) \, , 
\label{be1}\\
\varepsilon _+ (k,x) &= \varepsilon _3(k) I_1(x) + \varepsilon _4(k) K_1(x)\, , 
\label{be2}
\end{align}
 where $J_1$, $Y_1$ ($I_1$, $K_1$) denote ordinary (modified) Bessel functions of index $1$, while the coefficients $\varepsilon_j(k)$ (with  $j=1,\,2,\,3,\,4$) have to be assigned via the initial condition $\psi(0,r,z) = \psi _0(r,z) + \delta (r)$. In this scheme, taking into account that the $r$ variable is bounded and that $I_1(\infty)$ is divergent, the flux function $\psi(t,r,z)$ admits the following representation:
\begin{align}
\psi(t,r,z)&=-{A_0}/{A_1}+\Lambda r^{2}+\nonumber\\
&+e^{-\gamma t} \int_{k_0}^{k^*}\!\!\!\!\!dk\Big[\big[r\varepsilon_1(k)J_1(r\sqrt{|k^2-k^{*2}|}) \nonumber\\
&+r\varepsilon _2(k) Y_1(r\sqrt{|k^2-k^{*2}|})\big]e^{ikz}+ c.c.\Big] \nonumber\\
&+e^{-\gamma t}\int_{k^*}^{\infty}\!\!\!\!\!dk\Big[r\varepsilon_4(k) K_1(r\sqrt{|k^2-k^{*2}|})e^{ikz}+c.c.\Big]\,. 
\label{egsp35}
\end{align}
Here, $\Lambda=-\pi c^2C_1/A_1^2$ and, to exclude wave-lengths greater than the machine diameter, we have introduced a minimum wavenumber $k_0=\pi/a$, \ie $k\geqslant k_0$. We also remark that, by suitably choosing the constant $C_0$ in \eref{egn1} for the plasma pressure, the basic requirement $p\geqslant0$ can be easily implemented in our confined plasma region.

\subsection{Specific implementation}\label{sec3a}
In order to investigate the morphology of the plasma configuration, we analyze the level surfaces of $\psi(r,z,t)$ at given times, together with the surface $p=0$ (representing the plasma boundary layer). 
The general solution for $\psi$ as in \eref{egsp35} can be adapted to a given scenario by imposing specific initial conditions. In this respect, for the sake of simplicity, we assign to the functions $\varepsilon_j(k)$ a set of sufficiently narrow Gaussians, centered around arbitrarily given wave vectors $k_{j,i}$ and weighted by amplitudes $\bar{\varepsilon}_{j,i}$. 
Then, a given set of points $(r_l,z_l)$ lying along the boundary curve of the addressed scenario generates an associated set of algebraic equations of the form $\psi(r_l,z_l,0)=\psi_B$, where $\psi_B$ is the value of the magnetic flux at the plasma boundary. 
Since we require $p=0$ on the same surface, recalling \eref{egn1} and that $C_2=0$, we set $C_0=-C_1\psi_B$. The rest of the constants have to be determined according to the relevant plasma parameters.

As an illustrative example, let us assume the parameters characterizing a Tokamak equilibrium specified for the DTT facility, as in \citet{albanese19}. In particular, the main machine parameters are major radius $R_0=2.11\,$m, minor radius $a=0.64\,$m, averaged electron density $n_e=1.8\times10^{20}\,$m$^{-3}$ and electron temperature $T_e=6.1\,$keV. The resulting plasma frequency is $\omp=7.57\times10^{11}\,$s$^{-1}$ and, considering the Spitzer electric conductivity for a hydrogen plasma (with the Coulomb logarithm of $\mathcal{O}(10)$), we obtain $\sigma=5.34\times10^{8}\,\Omega^{-1}$m$^{-1}$. We implement an initial condition matching the double-null scenario, with main plasma parameters as reported in Table \ref{tab1}. In the same Table, we also report the fitted values associated to our analytic solution, which is able to correctly reproduce most of the parameters. 
In this scheme, we obtain a configuration characteristic time of $\gamma^{-1} = 1.1\times10^4\,$s. As a stability check, the safety factor $q$ meets the Kruskal-Shafranov condition for stability $q>1$ over the whole plasma region, with an average value of $2.6$.
\begin{figure}
\centering
\includegraphics[width=0.24\columnwidth]{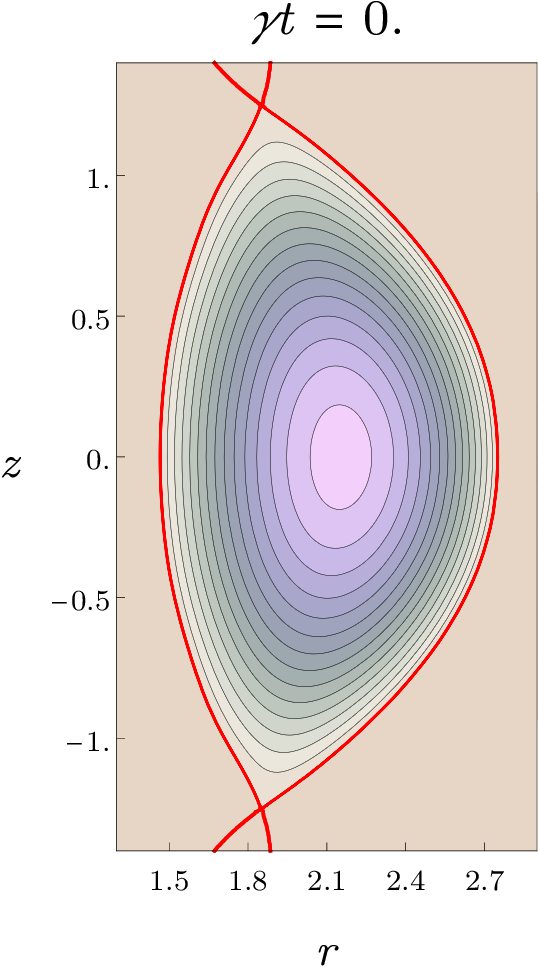}\hfill
\includegraphics[width=0.24\columnwidth]{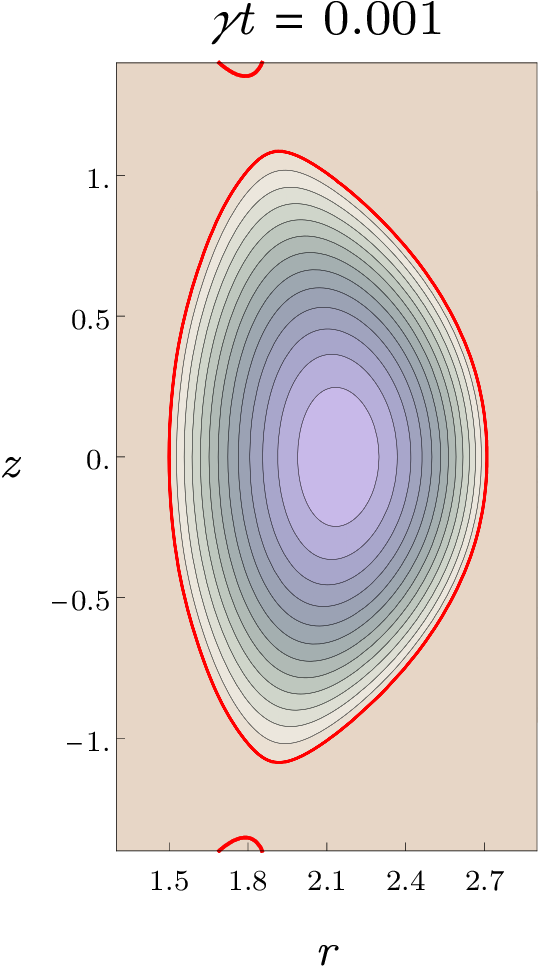}\hfill
\includegraphics[width=0.24\columnwidth]{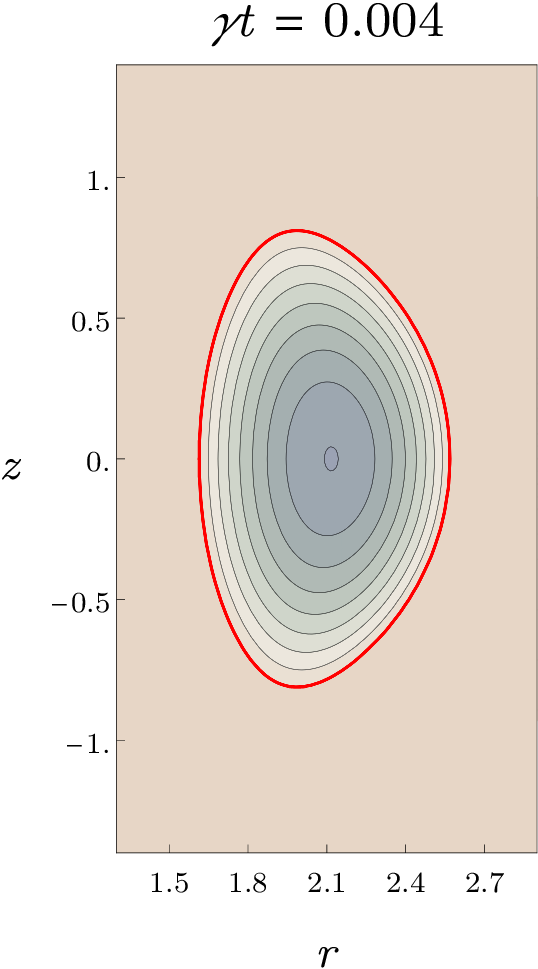}\hfill
\includegraphics[width=0.24\columnwidth]{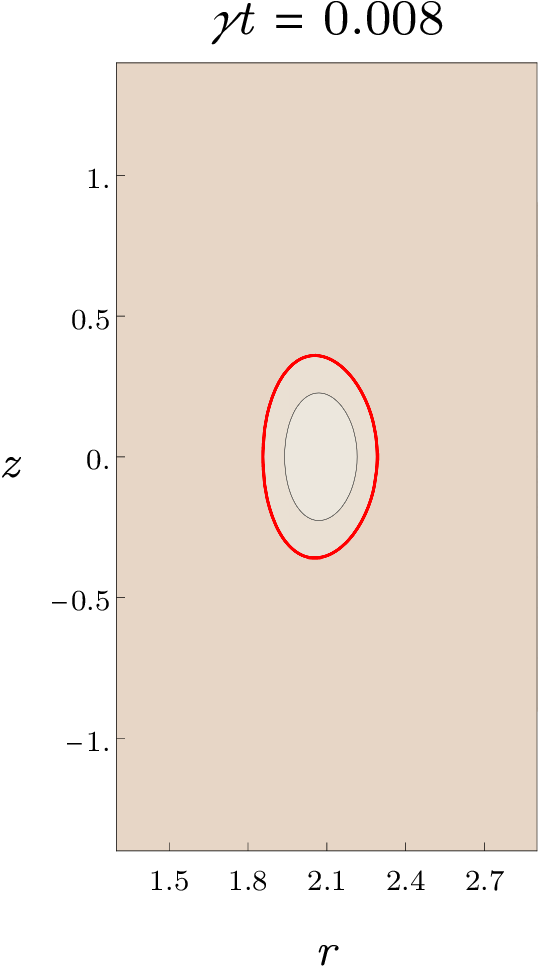}
\caption{
(Color online) Contour plot of the flux function (in the physical plane $(r,z)$) integrated from \eref{egsp35} according to the proposed double-null scenario for DTT \citep{albanese19}, for different instants as indicated over the panels (color scheme from beige ($\psi=2.50\,$Vs, also red line) to purple ($\psi=8\,$Vs)). The separatrix $p=0$ is enlightened in red. The initial condition is imposed through 14 boundary points, plus two conditions on the derivatives of $\psi$ at the x-point. Wavenumbers $k$ run from $0.52$ to $1.72$ in steps of $0.2$.
\label{fig_psi}}
\end{figure}

In \figref{fig_psi}, we plot the level surfaces of the flux function $\psi(t,r,z)$ at different times, where the initial condition at $t=0$ is shown in the first panel. At later stages, the allowed domain for the plasma configuration decreases, and the central pressure is correspondingly suppressed (cf. \eref{egn1}). Since the area inside the separatrix is decreasing in time, the axial symmetry implies that the confined plasma volume is also diminishing, but keeping a constant plasma density $\rho\equiv\rho_0$. As a consequence, the plasma evolution has to be associated with a loss of particles through the boundary layer of the toroidal plasma profile. The behaviour of this outgoing flux of matter must be described in a different physical setting, having to deal with the behaviour of non confined plasma in the scrape-off layer.

Concerning the lifetime of the configuration, it is important to stress that we observe the opening of all magnetic lines, determining the loss of confinement, at $t=99\,$s. This timescale is two orders of magnitude shorter than $\gamma^{-1}$, and is comparable with the predicted duration of the discharge of about $\simeq50\,$s. This behaviour can be understood if we consider that the plasma region can also be defined as the points satisfying $\psi\geqslant\psi_\text{B}$. Indicating the initial peak value of the magnetic flux as $\psi_\text{A}$, in correspondence to the magnetic axis, it is clear that after an overall decrease in $\psi$ of the order $\Delta\psi\equiv|\psi_\text{A}-\psi_\text{B}|$ the whole profile will lie below the $\psi_\text{B}$ threshold, \ie all magnetic lines will be open. Then, it is natural to define an effective lifetime according to the condition $\bar{\psi}_0 \left( 1-e^{-\gamma t^*}\right)=\Delta\psi$ (where $\bar{\psi}_0$ is the order of magnitude of the function $\psi_0(r,z)$), which provides the expression
\begin{equation}
t^*=-\gamma^{-1}\ln\left(1-\Delta\psi/\bar{\psi}_0\right)\,.
\label{eqtime}
\end{equation}
In the case under study, $\Delta\psi\simeq5.5\,$Vs and $\bar{\psi}_0\simeq600\,$Vs, so that $t^*=99\,$s, as shown in the plots.
{\begin{table}			
\centering
\caption{Data relative to the double-null DTT  scenario, as in \citet{albanese19}, and the corresponding fitted values from our solution and the Solov'ev configuration, which is introduced in Sec.\ref{sec4}. The subscript $\text{A}$ refers to quantities along the magnetic axis, $\psi_\text{B}$ is the magnetic flux at the plasma boundary, $\beta_\text{p}$ is the ratio of the plasma pressure to the poloidal magnetic pressure, $I_\text{P}$ is the total plasma current and $l_\text{i}$ is the internal inductance.}
\label{tab1}
\begin{tabularx}{\linewidth}{@{}lYYY@{}}
\toprule
   &  DTT\newline scenario & Compatible\newline configuration & Solov'ev\newline configuration \\
\midrule
$r_\text{A}\,$(m) & 2.17 & 2.16 & 2.15 \\
$\psi_\text{B}\,$(Vs) & 2.50 & 2.55 & 2.53 \\
$\psi_\text{A}\,$(Vs) & 11.48 & 8.02 & 7.94 \\
$B_\text{A}\,$(T) & 6.19 & 6.23 & 6.23 \\
$\beta_\text{p}$ & 0.43 & 0.43 & 0.43 \\
$I_\text{P}\,$(MA) & 5.00 & 5.00 & 5.00 \\
$l_\text{i}$ & 0.80 & 0.39 & 0.38 \\
\bottomrule
\end{tabularx}
\end{table}}

We also remark that the solution outside the boundary layer takes a different character, being described by a vacuum problem at the initial stage of the evolution. In this outer region the current density must be set to zero, according to the pressure profile. Therefore, outside the separatrix $p=0$, we must require that $A_1=A_0=C_1=C_0\equiv0$ and also that the toroidal current $J_{\phi}$ vanishes. This last condition leads to the equation
\begin{equation}
\Delta^* \psi(t,r,z)=0\;, 
\label{egszz}
\end{equation} 
which is the only surviving equation for the vacuum configuration. Clearly, the time dependence of the magnetic flux function in vacuum is ensured by the matching conditions on the boundary layer.

\subsection{Implications of temperature dynamics}
As already noted at the end of Sec.\ref{sec1}, the continuity equation \reff{egs14}, in the absence of velocity fields, implies a time independent profile of the mass density $\rho=\rho_0(r,z)$. In the limit of applicability of the perfect gas law to the plasma (coherent with a non-zero resistivity), we thus see that the temperature must decay in time like the pressure does, as implied by \erefs{egn1} and \reff{egs18}.

From this point of view, the assumption of dealing with a constant conductivity, on which the present analysis is based, is questionable. In fact, it is well known from \citet{Spitzer53} that the plasma conductivity increases as $\sigma \sim T^{3/2}$, therefore the validity of the present scheme extends as far as a time averaged value of this quantity can be considered, namely for a timescale $t \ll \gamma^{-1}$. 

The question could be addressed in a more rigorous way by including the temperature dynamics in the model, via the conservation equation for energy. The simple diffusive equation we could assign for the temperature evolution in cylindrical coordinates is the following:
\begin{equation}
\partial _tT = \frac{2}{3n_0K_B}\frac{1}{r}
\partial_r\left(r \kappa _T\partial _rT\right) + \frac{2}{3n_0K_B} 
\partial_z\left(\kappa _T \partial _zT\right)
\, , 
\label{xtempx}
\end{equation}
where $\kappa _T$ denotes the thermal conductivity coefficient, $K_B$ is the Boltzmann constant and $n_0(r,z)$ is the equilibrium plasma particle density, equal to $\rho_0(r,z)/m_i$ in the case of a single atomic species plasma with ion mass $m_i$. However, assuming the validity of the perfect gas law for the plasma, we must also have $T = p(\psi)/n_0$, which clearly opens a problem of compatibility between the equation above and \eref{egs18}, describing the flux function evolution.

This compatibility request cannot be easily solved and shows, once again, how the complete self-consistency of a plasma evolution implies serious restrictions on the relations among the involved physical quantities. Our analysis calls attention on how fixing the equilibria and then evolving transport on that configuration, recalculating it on a later step (and iterating the procedure, as in typical fusion codes), could cut off all these mathematical questions from the problem. In general, this procedure is reliable and efficient due to the different timescale of the equilibrium variation with respect to the transport phenomena, but it could lead to non-trivial miscalculations when the magnetic flux function and the other physical plasma quantities, for instance the velocity field, evolve together in a coherent and consistent MHD scheme. Our study of resistive diffusion acting on equilibria is just a simple example of a more general problem of consistency which could emerge when equilibrium and transport codes are matched together. 

Coming back to the question of the temperature behaviour, we observe that, as discussed in \citet{Montani18}, where an astrophysical context has been investigated, the compatibility of the system for $T=T(\psi)$ is still possible for a local model, where all the background quantities are considered constant, and the perturbations have a sufficiently short wavelength. A similar picture, with some minor modifications of the physical framework, could also be applied to a Tokamak configuration, if we were interested to study local effects of diffusion nearby a given regular magnetic surface $\psi(r,z) = \psi _0$. This study could be of interest in determining the stochastization of the magnetic flux dynamics and the onset of an island formation. This process is clearly absent in the present model, in which the magnetic flux surfaces are differentiable topological sets and evolve in time preserving this main feature and their basic shape. Using a local model nearby each assigned background surface, considering also a space-dependent resistivity, could allow a study of the instability of this scenario versus a stochastic domain.

Before closing this Section, it is worth recalling the hypotheses at the ground of the present analysis and their physical motivation or interpretation. We consider a non-steady (slowly varying) axisymmetric plasma configuration, characterized by zero velocity fields, a space-time independent temperature and a constant conductivity coefficient.

The assumption to deal with zero advective flows is rather natural in the analysis of equilibria \citep{biskamp} and it is justified by the absence of evidence for macroscopic plasma motion during a wide class of discharges in Tokamak devices. However, the observation of spontaneous toroidal and poloidal rotation, as well as rotation due to plasma interaction with hot neutral beams and other power sources \citep{Rice16} suggests that, under specific conditions, the present analysis needs to be extended in this direction. In the present work, the main task is to investigate the role of resistive diffusion in slowly altering an equilibrium configuration as time flows. Thus, we focus our attention on the resistive timescale as the fundamental one which drives the plasma evolution. In this sense, the introduction of a velocity field is surely of interest, but only after the basic resistive mechanism of diffusion is understood in its intrinsic nature.

The assumption of a uniform plasma temperature is more serious and definitely unrealistic close to the separatrix region of the plasma, but it is also commonly employed in the study of Tokamak physics, when the equilibrium configuration is viewed in its ground level features (see, e.g., \cite{tamain16}). More precisely, the possibility to consider a constant temperature must be implicitly thought as a restriction on the considered plasma space and time region. In the present case, this restriction must be referred to what we defined as the effective lifetime of the configuration, corresponding to the disappearance of a closed separatrix (see Sec. \ref{sec3a}). In fact, since this timescale is about two orders of magnitude shorter than the resistive diffusion time (derived in Sec.\ref{sec2}), it is safe to consider the plasma as isothermal during the evolution. As far as the plasma is sufficiently hot so that its collisional nature is negligible (\emph{e.g.}, the thermal conductivity discussed above), the isothermal nature of the plasma within the separatrix is a satisfactory approximation, up to the beginning of a disruption, before the thermal quench has occurred \citep{nedospasov08}.

For what concerns the constant nature of the conductivity coefficient, we can develop similar considerations about its restriction to a limited space and time region of the plasma, as argued for the temperature. However, in view of capturing the basic feature of a slowly varying equilibrium due to resistive diffusion, the uniformity of $\sigma$ seems to be natural. In fact, a non uniform coefficient $\sigma$ would only affect the diffusion properties in different space regions, without introducing any new conceptual modification of the present picture.

Although some of the above hypotheses could be weakened in a future semi-analytical analysis, we stress once more that the loss of an analytical treatment for a more general configuration outlines possible shortcomings in the complete separation between equilibria and transport, often introduced in plasma transport codes.


\section{Comparison with alternative approaches}\label{sec4}

In order to compare our self-consistent approach to other standard methods, we begin by studying an alternative analytical solution in correspondence to the well-known Solov'ev configuration \citep{solo68}. In this sense, we go back to the original GSE, \eref{egs13o}, and assume its right-hand side to be independent on $\psi$:
\begin{equation}
\Delta^*\psi= -16\pi^3C_{1s} r^2 -\frac{16\pi^2}{c^2}A_{1s} \,,
\label{esub1}
\end{equation}
where $C_{1s}$ and $A_{1s}$ are constants (the subscript $s$ indicates quantities relative to the Solov'ev scenario). The corresponding choices for $p(\psi)$ and $I(\psi)$ are the following:
\begin{equation}
p_s(\psi)=C_{1s}\psi+C_{0s}\,,\quad I_s(\psi)=\sqrt{2A_{1s}\psi+A_{0s}}\,,
\label{esub2}
\end{equation}
so that the pressure is of the same kind as previously considered (remember that $C_2=0$), while the axial current has a different functional form. It is important to remark that substituting the latter expression into \eref{egs-red1} we get
\begin{equation}
\frac{d^2I_s}{d\psi^2}|\nabla\psi|^2=-\frac{A_{1s}^2}{(2 A_{1s} \psi + A_{0s})^{3/2}}|\nabla\psi|^2=0\,
\end{equation}
which admits only the trivial solutions $A_{1s}=0$ or $\psi=const$. However, if the equation above is excluded from the model, \erefs{egs10a} and \eqref{esub1} lead to the expression
\begin{equation}
\psi_s(r,z,t)=- a(r)\, t + b(r) + \psi_0(r,z) \,,
\label{esub3}
\end{equation}
with
\begin{align}
a(r)&=\frac{4\pi^2 c^2}{\sigma}\left(C_{1s} r^2 +\frac{A_{1s}}{\pi c^2}\right)\,,\label{esv1}\\
b(r)&=2\pi^3 r^2\left[ -C_{1s} r^2 +\frac{2 A_{1s}}{\pi c^2} \left( 1-2\log r \right) \right]\,.\label{esv2}
\end{align}
Here, $\psi_0(r,z)$ is formally equivalent to the solution already considered in Sec.\ref{sec3}, since it must satisfy \erefs{egs22} and \eqref{egs23}, with the only difference that now $E_k=k^2$. The most striking feature of $\psi_s(r,z,t)$ is the linear time dependence, which differs from the exponential decay of the consistent solution. 

To test this discrepancy, we fit the new expression, \eref{esub3}, to the same double-null scenario of Sec.\ref{sec3a}. The agreement is sufficiently good, as can be noted from the fitted values in Table \ref{tab1}. Moreover, the time evolutions of the two profiles follow the same dynamics up to the loss of confinement, which, in this case, takes place after $98\,$s. This similarity between exponential and linear decays can be explained noting that confinement is lost on a timescale much shorter than $\gamma^{-1}$, when the exponential in \eref{egsp35} is still in its linear phase.

Although this result suggests that \eref{egs-red1} can be safely disregarded in the DTT plasma scenario, this cannot be considered as a general proof. The Solov'ev case has the good property of preserving the linearity of the system, which instead is usually broken in the context of numerical equilibrium solvers, such as EFIT \citep{lao85}, where nonlinear forms of $dp/d\psi$ and $IdI/d\psi$ are assumed.
In such scenarios, \eref{egs13o} cannot be solved through simple analytic means, so an exact comparison lies outside the scope of the present work. We propose an effective estimate of the incompatibility of the nonlinear case, by considering the following generalization of \eref{egs12}:
\begin{equation}
I_n(\psi)=A_{1,n}\psi^n+A_{0,n}\,,
\label{eqin}
\end{equation}
where the coefficients $A_{1,n}$ and $A_{0,n}$ are determined according to the relevant plasma parameters of Table \ref{tab1}. Assuming $\left|\nabla\psi\right|^2$ to be of the same order of magnitude in all configurations (\ie $\sim(\Delta\psi/a)^2$), the error committed in \eref{egs-red1} is quantified by the second derivative of $I_n$ with respect to $\psi$:
\begin{equation}
\frac{d^2I_n}{d\psi^2}=n(n-1) A_{1,n} \psi^{n-2}\,.
\label{eqi2n}
\end{equation}
The same quantity, calculated for the Solov'ev configuration, is taken as a reference, so we study the function
\begin{equation}
\epsilon(\psi,n)\equiv\log_{10}\left| \frac{d^2I_n}{d\psi^2}\middle/\frac{d^2I_s}{d\psi^2}\right|=\left|A^2_{1s}\frac{n(n-1) A_{1,n} \psi^{n-2}}{(2 A_{1s}\psi+A_{0s})^{3/2}}\right|\,,
\label{eqestim}
\end{equation}
defined as a logarithm for convenience.
\begin{figure}
\centering
\includegraphics[width=0.7\columnwidth]{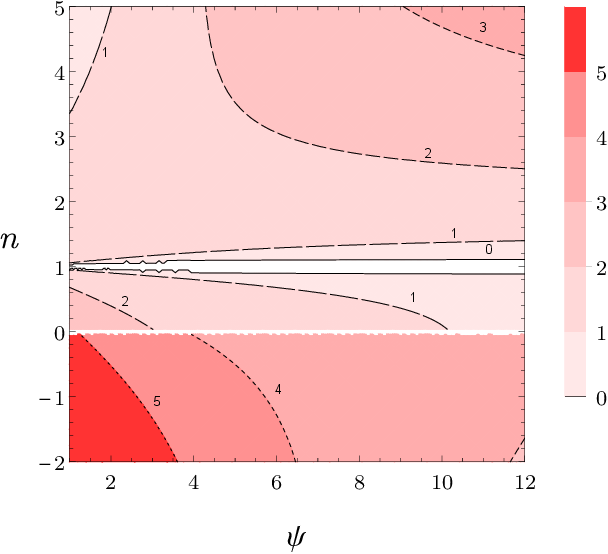}
\caption{(Color online) Variation of $\epsilon(\psi,n)$ for values of $\psi$ corresponding to the DTT double-null plasma scenario \citep{albanese19}, and $n\in(-2,5)$. The color scheme goes from white ($\epsilon<0$) to red ($\epsilon>5$), while each contour is labeled by the corresponding value, with shorter dashes indicating greater values.}
\label{fig_eps}
\end{figure}

In \figref{fig_eps}, it clearly emerges how the magnitude of $\epsilon$ grows quickly for $n$ different than 0 and 1 (the only two analytically correct values). In particular, the whole region below $n=0$ takes up values larger than 3, \ie the left-hand side of \eref{egs-red1} is at least three times larger in these cases than in the Solov'ev case. A fiducial interval can be defined around $n=1$, in which the discrepancy is less than one order of magnitude. According to this estimate, more detailed studies should be performed on the viability of the coupling of evolutive codes with nonlinear GSE configurations.

\section{Concluding remarks}\label{sec5}
We analyzed a varying tokamak plasma equilibrium, in which the magnetic field profile is damped by resistive effects. In such a dynamical scheme, the GSE is coupled with an evolutionary equation for the magnetic flux function dynamics, \ie the induction equation.

The main result has been the determination of a lifetime for the plasma confinement, here discussed in the particular case of an initial condition corresponding to the $5\,$MA double-null scenario for the DTT tokamak proposal, as in \citet{albanese19}. A secondary, effective lifetime also arises from the observation of the loss of magnetic confinement on a timescale much shorter than expected, and comparable with the duration of the discharge.

Clearly, the present analysis cannot be directly applied to the discharge evolution of a Tokamak machine. In fact, during a discharge, the current is governed by inductive processes associated to the time dependence of the current running in the magnetic field coils. Furthermore, when a flat-top configuration is reached, it is maintained with a steady profile also via different current drive mechanisms, such as radiofrequencies coupled to the plasma.

Our analysis is mainly aimed at providing some physical insight in view of the following delicate matter. The ideal Grad-Shafranov equation is usually assumed to describe the plasma equilibrium at any given time, while the generalized Ohm's law, on which transport computations are based, takes into account all the non-ideal processes involved in the construction of the steady current profile, including the resistive diffusion. Hence, from a rigorous mathematical point of view, the matching of such different pictures is inconsistent since the evolution of transport quantities would clearly influence the behaviour in time of the magnetic flux function, which instead is taken as fixed between different steps of an iterating procedure.
Moreover, even ideal effects like the presence of velocity fields pose similar questions, whenever particle transport is calculated over an underlying equilibrium obtained from a static version of the GSE.

The physical predictivity of this standard strategy in describing Tokamak physics relies on the different timescales of the transport processes and of the equilibrium variation, but this approximation clearly fails when the plasma configuration is subject to abrupt modifications, like during the L-H transition \citep{hm3} or disruptions.

For instance, we show that the compatibility of the dynamics requires the poloidal current function $I$ to have a linear dependence on the magnetic flux function. As discussed in Sec.\ref{sec4}, this constraint could produce significant deviations between the steady and the dynamical versions of the GSE, namely when non-linear contributions to $I(\psi)$ are considered, as it is often the case in numerical codes \citep{lao85}.

In other words, the present analysis cannot be directly applied to the operation of a Tokamak device, but it suggests a more careful understanding of the underlying physical assumptions on which different codes separately face different pieces of Tokamak physics. In large sized machines, having also large discharge durations, the consistency of the dynamics, raised here, could play a significant role.

For what concerns the application of our model to the DTT scenario, it is worth observing that our capability to analytically reproduce the plasma configuration is affected by a certain degree of approximation, since our solution lacks the sufficient number of parameters to constrain all the relevant plasma quantities. In this respect, the amount of parameters that can be fixed is naturally related to the linear prescription $I\propto\psi$, which is a remarkable conceptual implication of including resistivity into the magnetic flux function dynamics.

We also studied two cases where this prescription is not respected, \emph{de facto} disregarding \eref{egs-red1}. In the Solov'ev configuration, which keeps the system linear, no dramatic changes are observed on the profile, while in nonlinear cases we obtain numerical evidence of a larger discrepancy. We conclude that before saying a definitive word on the relevance of resistive diffusion in the equilibrium properties, a more systematic study on commonly used nonlinear plasma codes could be of interest.

\bibliographystyle{jpp}
\bibliography{egse-jpp}

\begin{thebibliography}{31}
\expandafter\ifx\csname natexlab\endcsname\relax\def\natexlab#1{#1}\fi
\def\au#1{#1} \def\ed#1{#1} \def\yr#1{#1}\def\at#1{#1}\def\jt#1{\textit{#1}}
  \def\bt#1{#1}\def\bvol#1{\textbf{#1}} \def\vol#1{#1} \def\pg#1{#1}
  \def\publ#1{#1}\def\arxiv#1{#1}\def\org#1{#1}\def\st#1{\textit{#1}}

\bibitem[Albanese {\em et~al.\/}(2019)Albanese, Crisanti, Martin, Martone,
  Pizzuto \& project~proposal contributors]{albanese19}
{\sc \au{Albanese, R.}, \au{Crisanti, F.}, \au{Martin, P.}, \au{Martone, R.},
  \au{Pizzuto, A.} \& \au{project~proposal contributors, DTT}} \yr{2019}
  \bt{Divertor tokamak test facility, interim design report}. {\em Tech.
  Rep.\/}.  \org{ENEA}.

\bibitem[Albanese \& Pizzuto(2017)]{albanese17}
{\sc \au{Albanese, R.} \& \au{Pizzuto, A.}} \yr{2017}  \at{The dtt proposal. a
  tokamak facility to address exhaust challenges for demo: Introduction and
  executive summary}.  \jt{Fusion Engineering and Design}  \bvol{122},  \pg{274
  -- 284}.

\bibitem[Alladio \& Crisanti(1986)]{ac86}
{\sc \au{Alladio, F.} \& \au{Crisanti, F.}} \yr{1986}  \at{Analysis of {MHD}
  equilibria by toroidal multipolar expansions}.  \jt{Nuclear Fusion}
  \bvol{26}~(9),  \pg{1143--1164}.

\bibitem[Alladio {\em et~al.\/}(2017)Alladio, Micozzi, Apruzzese, Boncagni,
  D'Arcangelo, Giovannozzi, Grosso, Iafrati, Lampasi, Maffia, Mancuso,
  Piergotti, Rocchi, Sibio, Tilia, Tudisco \& Zanza]{proto17}
{\sc \au{Alladio, Franco}, \au{Micozzi, P.}, \au{Apruzzese, G.M.},
  \au{Boncagni, L.}, \au{D'Arcangelo, Ocleto}, \au{Giovannozzi, Edmondo},
  \au{Grosso, L.A.}, \au{Iafrati, Matteo}, \au{Lampasi, Alessandro},
  \au{Maffia, G.}, \au{Mancuso, A.}, \au{Piergotti, V.}, \au{Rocchi, G.},
  \au{Sibio, A.}, \au{Tilia, B.}, \au{Tudisco, Onofrio} \& \au{Zanza, V.}}
  \yr{2017} The proto-sphera experiment, an innovative confinement scheme for
  fusion.

\bibitem[Biskamp(1993)]{biskamp}
{\sc \au{Biskamp, Dieter}} \yr{1993} {\em Nonlinear Magnetohydrodynamics\/}.
  {\em Cambridge Monographs on Plasma Physics\/} .  \publ{Cambridge University
  Press}.

\bibitem[{Dini} {\em et~al.\/}(2011){Dini}, {Baghdadi}, {Amrollahi} \&
  {Khorasani}]{dini}
{\sc \au{{Dini}, F.}, \au{{Baghdadi}, R.}, \au{{Amrollahi}, R.} \&
  \au{{Khorasani}, S.}} \yr{2011}  \at{{An Overview of Plasma Confinement in
  Toroidal Systems}}.  \jt{Horizons in World Physics}  \bvol{271},  \pg{71}.

\bibitem[{Grad}(1974)]{Grad:74}
{\sc \au{{Grad}, H.}} \at{ \yr{1974} } \jt{Advances in Plasma Physics}
  \bvol{5},  \pg{103}.

\bibitem[{Grad} \& {Hogan}(1970)]{Grad:70}
{\sc \au{{Grad}, Harold} \& \au{{Hogan}, John}} \yr{1970}  \at{{Classical
  Diffusion in a Tokomak}}.  \jt{Physical Review Letters}  \bvol{24}~(24),
  \pg{1337--1340}.

\bibitem[Grad \& Hu(1977)]{Grad:77}
{\sc \au{Grad, H.} \& \au{Hu, P.~N.}} \yr{1977}  \bt{Classical diffusion:
  theory and simulation codes}.  \org{{\em Tech. Rep.\/}}. United States,
  cONF-7709167--.

\bibitem[{Grad} {\em et~al.\/}(1975){Grad}, {Hu} \& {Stevens}]{Grad:75}
{\sc \au{{Grad}, H.}, \au{{Hu}, P.~N.} \& \au{{Stevens}, D.~C.}} \yr{1975}
  \at{{Adiabatic Evolution of Plasma Equilibrium}}.  \jt{Proceedings of the
  National Academy of Science}  \bvol{72}~(10),  \pg{3789--3793}.

\bibitem[{Grad} {\em et~al.\/}(1977){Grad}, {Hu}, {Stevens} \&
  {Turkel}]{Gradeta:77}
{\sc \au{{Grad}, H.}, \au{{Hu}, P.~N.}, \au{{Stevens}, D.~C.} \& \au{{Turkel},
  E.}} \yr{1977} {Classical plasma diffusion}.  \bt{In {\em Plasma Physics and
  Controlled Nuclear Fusion Research 1976, Volume 2\/}}, ,  \vol{vol.~2},
  \pg{pp. 355--365}.

\bibitem[{Grad} \& {Rubin}(1958)]{grad}
{\sc \au{{Grad}, H.} \& \au{{Rubin}, H.}} \yr{1958}  \at{{Hydromagnetic
  equilibria and force-free fields}}.  \jt{Proc. 2nd United Nations Conference
  Peaceful Use At. Energy}  \bvol{31},  \pg{190}.

\bibitem[{Keilhacker}(1987)]{hm2}
{\sc \au{{Keilhacker}, M.}} \yr{1987}  \at{{H-mode confinement in tokamaks}}.
  \jt{Plasma Physics and Controlled Fusion}  \bvol{29}~(10A),  \pg{1401--1413}.

\bibitem[Landau \& Lifshitz(1984)]{landau8}
{\sc \au{Landau, L.D.} \& \au{Lifshitz, E.M.}} \yr{1984}  \at{Chapter viii -
  magnetohydrodynamics}. In {\em Electrodynamics of Continuous Media (Second
  Edition)\/},  \bt{Second edition edn. (ed. \ed{L.D. Landau \& E.M.
  Lifshitz})},  \st{Course of Theoretical Physics},  \vol{vol.~8},  \pg{pp. 225
  -- 256}.  \publ{Amsterdam: Pergamon}.

\bibitem[Lao {\em et~al.\/}(1985)Lao, John, Stambaugh, Kellman \&
  Pfeiffer]{lao85}
{\sc \au{Lao, L.L.}, \au{John, H.~St.}, \au{Stambaugh, R.D.}, \au{Kellman,
  A.G.} \& \au{Pfeiffer, W.}} \yr{1985}  \at{Reconstruction of current profile
  parameters and plasma shapes in tokamaks}.  \jt{Nuclear Fusion}
  \bvol{25}~(11),  \pg{1611--1622}.

\bibitem[{Mc Carthy}(1999)]{mccar99}
{\sc \au{{Mc Carthy}, P.~J.}} \yr{1999}  \at{{Analytical solutions to the
  Grad-Shafranov equation for tokamak equilibrium with dissimilar source
  functions}}.  \jt{Physics of Plasmas}  \bvol{6}~(9),  \pg{3554--3560}.

\bibitem[{Miller}(1985)]{Miller:85}
{\sc \au{{Miller}, G.}} \yr{1985}  \at{{Resistive evolution of general plasma
  configurations}}.  \jt{Physics of Fluids}  \bvol{28}~(5),  \pg{1354--1358}.

\bibitem[{Montani} {\em et~al.\/}(2018){Montani}, {Rizzo} \&
  {Carlevaro}]{Montani18}
{\sc \au{{Montani}, Giovanni}, \au{{Rizzo}, Mariachiara} \& \au{{Carlevaro},
  Nakia}} \yr{2018}  \at{{Behavior of thin disk crystalline morphology in the
  presence of corrections to ideal magnetohydrodynamics}}.  \jt{Physical Review
  E}  \bvol{97}~(2),  \pg{023205},  \arxiv{arXiv: 1802.10506}.

\bibitem[Nedospasov(2008)]{nedospasov08}
{\sc \au{Nedospasov, A.V.}} \yr{2008}  \at{Thermal quench in tokamaks}.
  \jt{Nuclear Fusion}  \bvol{48}~(3),  \pg{032002}.

\bibitem[Nührenberg(1972)]{Nuhrenberg:72}
{\sc \au{Nührenberg, J.}} \yr{1972}  \at{Special time-dependent solutions of
  diffuse tokamak equilibrium}.  \jt{Nuclear Fusion}  \bvol{12}~(2),
  \pg{157--163}.

\bibitem[{Pao}(1976)]{Pao:76}
{\sc \au{{Pao}, Y.~P.}} \yr{1976}  \at{{Classical diffusion in toroidal
  plasmas}}.  \jt{Physics of Fluids}  \bvol{19}~(8),  \pg{1177--1182}.

\bibitem[{Reid} \& {Laing}(1979)]{Reid:79}
{\sc \au{{Reid}, J.} \& \au{{Laing}, E.~W.}} \yr{1979}  \at{{The resistive
  evolution of force-free magnetic fields. Part 1. Slab geometry}}.
  \jt{Journal of Plasma Physics}  \bvol{21}~(3),  \pg{501--510}.

\bibitem[Rice(2016)]{Rice16}
{\sc \au{Rice, J.E.}} \yr{2016}  \at{Experimental observations of driven and
  intrinsic rotation in tokamak plasmas}.  \jt{Plasma Physics and Controlled
  Fusion}  \bvol{58}~(8),  \pg{083001}.

\bibitem[{Shafranov}(1966)]{shafranov}
{\sc \au{{Shafranov}, V.~D.}} \yr{1966}  \at{{Plasma Equilibrium in a Magnetic
  Field}}.  \jt{Reviews of Plasma Physics}  \bvol{2},  \pg{103}.

\bibitem[{Solov'ev}(1968)]{solo68}
{\sc \au{{Solov'ev}, L.~S.}} \yr{1968}  \at{{The Theory of Hydromagnetic
  Stability of Toroidal Plasma Configurations}}.  \jt{Soviet Journal of
  Experimental and Theoretical Physics}  \bvol{26},  \pg{400}.

\bibitem[{Spitzer} \& {H{\"a}rm}(1953)]{Spitzer53}
{\sc \au{{Spitzer}, Lyman} \& \au{{H{\"a}rm}, Richard}} \yr{1953}
  \at{{Transport Phenomena in a Completely Ionized Gas}}.  \jt{Physical Review}
   \bvol{89}~(5),  \pg{977--981}.

\bibitem[{Strand} \& {Houlberg}(2001)]{strand01}
{\sc \au{{Strand}, P.I.} \& \au{{Houlberg}, W.A.}} \yr{2001}  \at{{Magnetic
  flux evolution in highly shaped plasmas}}.  \jt{Physics of Plasmas}
  \bvol{8}~(6),  \pg{2782--2792}.

\bibitem[Tamain {\em et~al.\/}(2016)Tamain, Bufferand, Ciraolo, Colin, Galassi,
  Ghendrih, Schwander \& Serre]{tamain16}
{\sc \au{Tamain, P.}, \au{Bufferand, H.}, \au{Ciraolo, G.}, \au{Colin, C.},
  \au{Galassi, D.}, \au{Ghendrih, Ph.}, \au{Schwander, F.} \& \au{Serre, E.}}
  \yr{2016}  \at{The tokam3x code for edge turbulence fluid simulations of
  tokamak plasmas in versatile magnetic geometries}.  \jt{Journal of
  Computational Physics}  \bvol{321},  \pg{606--623}.

\bibitem[{Wagner}(2007)]{hm3}
{\sc \au{{Wagner}, F.}} \yr{2007}  \at{{A quarter-century of H-mode studies}}.
  \jt{Plasma Physics and Controlled Fusion}  \bvol{49}~(12B),  \pg{B1--B33}.

\bibitem[{Wagner} {\em et~al.\/}(1984){Wagner}, {Fussmann}, {Grave},
  {Keilhacker}, {Kornherr}, {Lackner}, {McCormick}, {M{\"u}ller},
  {St{\"a}bler}, {Becker}, {Bernhardi}, {Ditte}, {Eberhagen}, {Gehre},
  {Gernhardt}, {Gierke}, {Glock}, {Gruber}, {Haas}, {Hesse}, {Janeschitz},
  {Karger}, {Kissel}, {Kl{\"u}ber}, {Lisitano}, {Mayer}, {Meisel}, {Mertens},
  {Murmann}, {Poschenrieder}, {Rapp}, {R{\"o}hr}, {Ryter}, {Schneider},
  {Siller}, {Smeulders}, {S{\"o}ldner}, {Speth}, {Steuer}, {Szymanski} \&
  {Vollmer}]{hm1}
{\sc \au{{Wagner}, F.}, \au{{Fussmann}, G.}, \au{{Grave}, T.},
  \au{{Keilhacker}, M.}, \au{{Kornherr}, M.}, \au{{Lackner}, K.},
  \au{{McCormick}, K.}, \au{{M{\"u}ller}, E.~R.}, \au{{St{\"a}bler}, A.},
  \au{{Becker}, G.}, \au{{Bernhardi}, K.}, \au{{Ditte}, U.}, \au{{Eberhagen},
  A.}, \au{{Gehre}, O.}, \au{{Gernhardt}, J.}, \au{{Gierke}, G.~V.},
  \au{{Glock}, E.}, \au{{Gruber}, O.}, \au{{Haas}, G.}, \au{{Hesse}, M.},
  \au{{Janeschitz}, G.}, \au{{Karger}, F.}, \au{{Kissel}, S.},
  \au{{Kl{\"u}ber}, O.}, \au{{Lisitano}, G.}, \au{{Mayer}, H.~M.},
  \au{{Meisel}, D.}, \au{{Mertens}, V.}, \au{{Murmann}, H.},
  \au{{Poschenrieder}, W.}, \au{{Rapp}, H.}, \au{{R{\"o}hr}, H.}, \au{{Ryter},
  F.}, \au{{Schneider}, F.}, \au{{Siller}, G.}, \au{{Smeulders}, P.},
  \au{{S{\"o}ldner}, F.}, \au{{Speth}, E.}, \au{{Steuer}, K.~H.},
  \au{{Szymanski}, Z.} \& \au{{Vollmer}, O.}} \yr{1984}  \at{{Development of an
  Edge Transport Barrier at the H-Mode Transition of ASDEX}}.  \jt{Physical
  Review Letters}  \bvol{53}~(15),  \pg{1453--1456}.

\bibitem[Wesson(2011)]{wesson}
{\sc \au{Wesson, John}} \yr{2011} {\em Tokamaks, 4th Edition\/}. {\em
  International Series of Monographs in Physics\/} .  \publ{Oxford Science
  Publications}.

\end{thebibliography}

\end{document}